\documentclass[preprint,12pt]{elsarticle}
\usepackage[latin1]{inputenc}

\usepackage{epsfig}

\usepackage{amssymb}
\usepackage{amsthm}
\usepackage{amsmath}
\usepackage[caption=false,font=footnotesize]{subfig}

\usepackage{url}
\usepackage{array}
\usepackage{booktabs}
\usepackage{graphicx}
\usepackage{makecell}
\usepackage{lipsum}
\usepackage{cases}
\newcommand{\fig}[1]{Figure~\ref{#1}}

\journal{Applied Energy}

\begin{document}

\begin{frontmatter}



\title{Modeling Time-dependent CO$_2$ Intensities in Multi-modal Energy Systems with Storage\tnoteref{t1}}
\tnotetext[t1]{This research did not receive any specific grant from funding agencies in the public, commercial, or not-for-profit sectors.}

\author{Christopher Ripp\corref{cor1}}
\cortext[cor1]{Corresponding author. Department
of Electrical Engineering and Information Technology, Technical University Darmstadt, Darmstadt, Germany}
\ead{christopher.ripp@eins.tu-darmstadt.de}
\author{Florian Steinke}
\address{Department
	of Electrical Engineering and Information Technology, Technical University Darmstadt, Darmstadt, Germany}

\begin{abstract}
CO$_2$ emission reduction and increasing volatile renewable energy generation mandate stronger energy sector coupling and the use of energy storage.
In such multi-modal energy systems, 
it is challenging to determine the effect of an individual player's consumption pattern onto overall CO$_2$ emissions.
This, however, is often important to evaluate the suitability of local CO$_2$ reduction measures. 
Due to renewables' volatility, the traditional approach of using annual average CO$_2$ intensities per energy form is no longer accurate, but the time of consumption should be considered. 
Moreover, CO$_2$ intensities are highly coupled over time and different energy forms due to sector coupling and energy storage.
We introduce and compare two novel methods for computing time-dependent CO$_2$ intensities, that address different objectives:
the first method determines CO$_2$ intensities of the energy system as is.
The second method analyzes how overall CO$_2$ emissions would change in response to infinitesimal demand changes.
Given a digital twin of the energy system in form of a linear program, we show how to compute these sensitivities very efficiently.
We present the results of both methods for two simulated test energy systems
and discuss their different implications. 
\end{abstract}


\end{frontmatter}


\section{Introduction}

The Paris climate agreement mandates a drastic decrease of CO$_2$ emissions in order to keep climate change within the two degree goal for global warming. 
Besides various efficiency improvements in energy production and consumption technologies,
the energy sector needs to accommodate a very high share of renewable electricity to reach this goal \citep{EU11}. 
This development is already under way, with member states of the EU predicted to hit their renewable electricity target of 20\% for 2020, further aiming for 27\% by 2030 \citep{eurostatrenewable2017}. 
From a technical point of view, even 100\% renewable electricity might be technically realizable \citep{lund2007renewable}. 
However, such systems require more flexible energy consumption and the interlinking of all energy sectors to balance volatile renewable production 
-- a trend often called sector coupling \citep{Sector_Coupling_Steinke}, multimodal or multi-energy systems \citep{MANCARELLA20141}.

During the course of this energy transition, any player in the system has to take numerous decisions about how to change his local behavior and energy setup. 
For example, a university has to decide whether to invest in combined heat and power plants on campus or in electric heating with heat pumps, how far to extend its internal heat distribution grid, and how to operate all of these components together in an optimal way.
Besides the financials, the CO$_2$ effect of any such local action needs to be well-understood.
It could also be the basis for CO$_2$ taxes or levies to incentivize emission-reducing behavior.
Since individual local energy changes do not change the system as a whole a lot, a straight-forward way for estimating the CO$_2$ effect of local actions is to use the CO$_2$ intensities of all energy forms consumed from the enregy system and multiply these with the projected changes of demand or generation patterns.
Computing CO$_2$ intensities such that they yield the plausible implications is, however, not a trivial task.

Several known approaches use annual average CO$_2$ intensities, see section~\ref{sec_related}.
Since highly volatile renewables impact the electricity generation mix in a strongly time-dependent manner, the CO$_2$ intensity should, however, rather be regarded as a time-dependent quantity. 
None of the previous research does take into account the effect of large-scale storage or full-scale sector coupling, that is reasonable to assume for highly renewable energy systems.
Storage may be charged with CO$_2$ neutral energy, e.g. from solar power, in which case any output from that storage should also be regarded as CO$_2$ free.
If, however, it is charged with energy from fossil sources, this is not the case.
Sector coupling interlinks the CO$_2$ intensity from one energy form with the other, potentially with circular reasoning.
For example, a gas power plant may be regarded as producing CO$_2$ neutral electricity and heat if it is run on renewably generated gas, a fact that again depends on the CO$_2$ intensity of the electricity the gas was generated from. 

In this paper, we 
describe and compare two new methods to determine time-dependent CO$_2$ intensities.
Both methods are consistent for fully multi-modal energy systems and the use of significant storage.
The key idea of the first method, from now on called "as-is" analysis, is to exactly trace all CO$_2$ emissions related to any kind of energy flow or storage content throughout the whole energy system.
All energy forms are treated equally. 
The second method, from now on called "what-if" analysis, calculates the differential change in the total CO$_2$ emissions of the system with respect to an infinitesimal change of the demand in a given energy form and a given time.
As is shown in the experiments and discussed thereafter, 
the resulting CO$_2$ intensity values for the two approaches are often very different, as are the implications one should draw from them.
Depending on the exact purpose of the query, however, both approaches can suitably be employed.
%

Our work extends previous work, outlined in section~\ref{sec_related}, by correctly considering fully multi-modal systems including a solution to circular reasoning problems and as well as non-negligible storage.
Concerning a computational perspective, we present another important contribution: we show how to compute the CO$_2$ sensitivities of the what-if analysis very efficiently, at least for common linear programming (LP) optimization based energy system models.
Moreover, we compare two approaches for computing CO$_2$ intensities on concrete examples and discuss the meaning and application of each computational approach. 

The proposed methods are described in section~\ref{sec_methods}.
The computational trick for efficient what-if analysis is presented in section~\ref{sec_KKT}.
In section~\ref{sec_systems}, we demonstrate the results for both approaches for two test energy systems, 
highlighting the effects of large-scale storage as well as sector coupling onto time-dependent CO$_2$ intensities. 
In section \ref{sec_conc} summarize our findings, discuss the different benefits and limitations of our two approaches, and give a brief outlook.

\section{Related Work}\label{sec_related}

The evaluation and discussion of CO$_2$ emissions of various energy production technologies and the influences of changes in an electrical energy system on these emissions has been studied for some years (e.g. \citep{HAWKES20105977,HAWKES2014197,marnay2002estimating,VOORSPOOLS20001119,VOORSPOOLS2000967,REES20141220}). 
Depending on the scope of the studies, e.g. countries or technologies, different methodologies were used and different results were presented.
Yet there is no general rule which technology reduces the emissions at most and wont will be.
The effect of system changes on CO$_2$ has to be individually answered for each system design.
The most commonly used and widely accepted approach to  determine the effect on CO$_2$ emissions is the computation of the marginal CO$_2$ emission of the respectively regarded energy system.
Therefor, the marginal CO$_2$ intensity follows as the ratio between additional CO$_2$ divided by the additional use of energy.
The most common approaches to compute marginal CO$_2$ intensities used modeled or observed load duration curves in combination with merit orders to identify the marginal production source. 
Voorspools et. al presented in the year 2000 a new tool and methodolgy for the economic dispatch to estimate marginal and incremental emissions factors to point out the optimum measures to mitigate CO2 emissions for the electricity generation \citep{VOORSPOOLS20001119}.
In 2002 Marnay et. al calculated the historic marginal CO$_2$ emissions of the California electric power sector by using a load curve approach and compared this values to the average CO$_2$ emissions \citep{marnay2002estimating}. 
In a further research Voorspool et. al used this methodolgy to compute the difference of GHG emission between different predicted scenarios of the evolution of the belgium energy system \citep{VOORSPOOLS2000967}.
In 2010 Hawkes et. al introduced a new approach to calculate the so called short term marginal CO$_2$ intensity, which represents only little structural change, via linear regression  based on historic data of great britain' electric energy production from 2002-2009 \citep{HAWKES20105977}. 
They also compared the calculated historic average CO$_2$ intensities with the calculated marginal CO$_2$ intensities and made predictions of future behavior by slightly changing the underlying production mix via manipulating the historical data.
In 2014 they extend their work and introduced the so called long run marginal CO$_2$ intensity, which displays the effect of structural change in the electricity system \citep{HAWKES2014197}. 
Based on Hawkes linear regression approach Siler-Evans et. al computed various marginal green house gas emission intensities of the U.S. electricity system \citep{margCO2USA} based on historic data from 2006 through 2011.

All previous studies have in common that their focus lies on the estimation of CO$_2$ intensities of historic energy mix based on historical given data.
As well as they limit these estimations to the electric energy system and not taken into account that high potential in saving CO$_2$ lies in the conversion of power to x in higly RES pentrated systems.
Some studies, e.g. \citep{HAWKES2014197}, trying to adapt the historical system data to predict the effects on CO$_2$ emissions due to the manipulation related system changes.
None of these studies including the effects of short or specially long term storage technologies, which completely decouple the production from usage time, which strongly effects the CO$_2$ intensities of the used energy within a consumption-based accounting of CO$_2$. 
Another related and not yet discussed topic is the computation of marginal CO$_2$ intensities in LP economic dispatch models in terms of a sensitivity analysis.
Our proposed methods are able to consider the effect of storage on the CO$_2$ intensities and are able to compute the marginal CO$_2$ intensities for multi-energy systems as an effect of differential changes of the energy demand without the need of a new optimization.

The authors of \citep{graus2011methods,ANG201656} calculate and analyze annual average CO$_2$ intensities for the joint heat and electricity production. 
Based on these numbers, the effectiveness of different CO$_2$ reduction measures are discussed in \citep{HARMSEN2013803}.
Similarly, the German building efficiency directive accounts for energy purchases from the grid via fixed primary energy factors \citep{EnEv17} or the 
EU requires that electricity utilities report their annual CO$_2$ intensity on each bill \citep{EC_DIRECTIVE_2009}.
Since highly volatile renewables impact the electricity generation mix in a strongly time-dependent manner, the CO$_2$ intensity should, however, rather be regarded as a time-dependent quantity.
The effect of including such dynamic CO$_2$ intensities for electricity into demand response tariffs is examined in \citep{STOLL2014490}. 
Yet all of this previous research does not take into account the effect of large-scale storage or sector coupling.
Storage may be charged with CO$_2$ neutral energy, e.g. from solar power, in which case any output from that storage should also be regarded as CO$_2$ free.
If, however, it is charged with energy from fossil sources, this is not the case.
Sector coupling interlinks the CO$_2$ intensity from one energy form with the other.
For example, a gas power plant may be regarded as producing CO$_2$ neutral electricity and heat if it is run on renewably generated gas.
When trying to correctly account for the fuel consumption of combined-heat-and-power plants (CHP), 
\citep{mauch2010allokationsmethoden} compares the consumption of the joint system with that of reference technologies for separately producing electricity and heat, distributing the joint savings evenly among the different modalities. 
In a fully multi-modal energy system, however, this approach becomes difficult. 
First, many reference technologies are required. 
Second, even evaluating the CO$_2$ intensity of a single technology may be difficult, if for example the consumed gas is partially produced via power-to-gas, resulting in circular reasoning. 

\section{Methods for computing CO$_2$ intensities}\label{sec_methods}

\begin{figure}[t]
	\centering
	\includegraphics[scale=0.5]{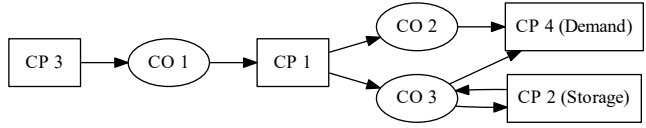}
	\caption{Schematic illustration of our abstraction of multi-modal energy systems. Ellipses encode energy forms (commodities) and rectangles energy conversion processes. Arrows indicate energy flows. }
	\label{general_model}
\end{figure}

For representing multi-modal energy systems we use the following modeling abstractions, see also \fig{general_model}: 
The different energy forms are called commodities (denoted as $co$, $co'$). 
At each point in time $t$, a conversion process ($cp$) takes amount $E_{in}(cp,co,t)$ of energy from input commodity $co$ and produces energy $E_{out}(cp,co',t)$ of output commodity $co'$.
Conversion processes can have several input and output commodities. 
Standard conversion processes, such as CP1 in \fig{general_model}, are augmented by storage processes, such as CP2, whose in and outflows of energy are buffered through an internal storage with time-dependent storage level $SL(cp,t)$. 
The model is completed by demand and import processes, see CP4 and CP3, respectively. 
These processes with only input or output energy flows constitute the model boundary, accounting for final energy use outside the model or the import of energies into the model domain. 

\subsection{As-is analysis}\label{sub_sec_as-is}
We now present the first of the two new methods for computing time-dependent specific CO$_2$ intensities $I(co,t)$ for each commodity $co$ and each time $t$.
Under this term we understand that fraction of the total energy system CO$_2$ emissions that is attributable to the generation of energy form $co$ at time $t$, divided by the total amount of such energy.

We track the absolute amount of attributable CO$_2$ emissions with each energy flow in and out of any conversion process. 
CO$_2$ outflows $M_{out}(cp,co,t)$ of a conversion process $cp$ are accounted for separately for each target commodity $co$, inflows $M_{in}(cp,t)$ are aggregated for the whole conversion process.
Time-dependent CO$_2$ intensities are then given as 
\begin{subequations}
	\label{eqn_asis}
	\begin{align}
	I^{as-is}(co,t) &= \frac{\sum_{cp} M_{out}(cp,co,t)}{\sum_{cp} E_{out}(cp,co,t)}.\label{equ_I}
	\end{align}
	Additionally, we track the CO$_2$ emissions $M_{stor}(cp,t)$ attributable to the generation of storage content $SL(cp,t)$ over time. This yields the current storage CO$_2$ intensity as 
	\begin{align}
	I_{stor}^{as-is}(cp,t) &=  \frac{M_{stor}(cp,t)}{SL(cp,t)}.\label{eqn_I_stor}
	\end{align}
	Numeric difficulties arise when the denominators are close to zero. We therefore introduce small minimum thresholds.
	\bigskip
	
	To obtain a consistent picture for the whole multi-modal energy system with storage, we now define a set of further linear equations for each type of conversion process.
	
	For standard conversion processes without storage we use equations
	\begin{align}
	M_{in}(cp,t) &= \sum_{co} E_{in}(co,cp,t) I^{as-is}(co,t), \label{equ_M_in}\\
	M_{out}(cp,co,t) &= M_{in}(cp,t) \frac{E_{out}(cp,co,t)}{\sum_{co'} E_{out}(cp,co',t)}.\label{equ_M_out}
	\end{align}
	$M_{in}(cp,t)$ aggregates the CO$_2$ emissions attributable to the generation of all energies consumed by the conversion process.
	This total CO$_2$ amount is then split evenly among all produced energy outputs, relative to the produced energy amount.
	Note that this equal treatment of all outputs is a key assumption of our approach.
	It may contradict common feelings and exergy arguments.
	Consider, for example, a CHP plant where one often regards electricity to be the real product and heat as only a side-effect, and thus CO$_2$ free.
	However, we are convinced that in complex multi-modal energy systems with high shares of renewables these uneven a priori valuations of different energy forms are no longer valid.
	CHP plants may often run more for their heat than for the electricity produced, if the wind is blowing. 
	Moreover, the equal treatment of all energy forms improves the explainability and interpretability of the approach.
	
	A second considered group of conversion processes are storage processes. To describe these, we use Eq.~\eqref{equ_M_in} and 
	\begin{align}
	M_{out}(cp,co,t) &= I_{stor}^{as-is}(cp,t-1) E_{out}(cp,co,t)/\eta_{stor}(cp), \label{eqn_Mstor_out}\\
	M_{stor}(cp,t) &= M_{stor}(cp,t-1) + M_{in}(cp,t) \notag\\
	&- \sum_{co} M_{out}(cp,co,t) \label{eqn_M_stor}.
	\end{align}
	The CO$_2$ output is given as the storage CO$_2$ intensity at the beginning of the time step multiplied by the energy taken out of storage during the time step. 
	The energy taken from storage is the the energy delivered to the product commodity divided by the storage efficiency $\eta_{stor}(cp)$.
	Eq.~\eqref{eqn_M_stor} encodes the time consistency of the CO$_2$ amount attributed to the storage energy content.
	
	The last group of conversion processes considers the system boundary.
	At each import process that takes energy into the model domain, e.g. for importing fossil fuels, we assign the related CO$_2$ content as 
	\begin{align}
	M_{out}(cp,co,t) &= E_{out}(cp,co,t)\eta_{CO2}(cp,co).
	\end{align}
\end{subequations}
We assume that all CO$_2$ imported into the model will eventually end up in the atmosphere.
Thus, $\eta_{CO2}(cp,co)$ encodes the CO$_2$ emissions of the fuel relative to its energetic value.
No equations are needed for demand processes.

The resulting set of linear equations~\eqref{eqn_asis} is solved simultaneously, yielding the CO$_2$ intensities of each commodity and storage within the multi-modal system for each point in time. 

\subsection{What-if analysis}\label{sub_sec_what-if}
Our second novel method aims at analyzing how the system would change, given infinitesimally small changes of the demand of commodity $co$ at time $t$.
Unlike the previous approach that worked with observed operational data only, 
we now need a model that allows determining the hypothetical system operation given modified demands. 
Such a model can be called a \emph{digital twin} of the energy system.\footnote{\url{https://en.wikipedia.org/wiki/Digital_twin}}

There are many ways to obtain digital twins of the energy system \citep{HerTorRei12}.
Our proposed approach for determining CO$_2$ intensities can work with all of them. 
For the exposition, however, we will focus on fundamental bottom up models, specifically partial equilibrium models \citep{SCHABER2012123,osemosys}, whose parameters are fitted to explain real behavior, see e.g. \citep{SCHABER2012123}. 

\bigskip
In the notation of this paper such partial equilibrium models can be sketched as follows.
We optimize over all energy flows $E_{in}(co,cp,t)$ and $E_{out}(cp,co,t)$ the cost of operation given as 
\begin{subequations}
	\begin{align}
	\sum_{cp,co,t} c_{var}(cp,co) E_{out}(cp,co,t),
	\end{align}
	where $c_{var}(cp,co)$ are the specific variable costs related to the energy flow $E_{out}(cp,co,t)$.
	The optimization is subject to a host of constraints.
	One is the conversion process-internal energy balance for each time step, 
	\begin{align}
	E_{out}(cp,co,t) = \kappa(cp,co) \sum_{co'} E_{in}(co',cp,t). \label{eqn_bottomupmodel_frac}
	\end{align}
	Here, $\kappa(cp,co)$ is the efficiency for output $co$. 
	Another constraint is the model-wide energy balance for each energy form and time,
	\begin{align}
	\sum_{cp} E_{out}(cp,co,t) = \sum_{cp'} E_{in}(co,cp',t). \label{eqn_energy_balance}
	\end{align}
	Moreover, outputs are capacity limited,
	\begin{align}
	0 \leq E_{out}(cp,co,t) \leq Cap(cp,co)
	\end{align}
	and some of them, namely the demands, are fixed to predefined levels,
	\begin{align}
	E_{in}(co,cp,t) = D(co,t).
	\end{align}
	This allows to compute total system-wide CO$_2$ emissions as 
	\begin{align}
	M_{tot} = \sum_{t,cp,co} \eta_{CO2}(cp,co) E_{out}(cp,co,t), \label{eqn_CO2_balance}
	\end{align}
	\label{eqn_opt_mod}
\end{subequations}
where $\eta_{CO2}(cp,co)$ are the CO$_2$ emissions that are produced by this conversion step and were not accounted before.
It usually suffices to assign the specific CO$_2$ emissions and fuel costs only to the import processes, that move energy into the model.
A full energy system model will typically contain many additional equations, e.g. to model storage processes or renewables with intermittent availability, see \citep{SCHABER2012123,osemosys}.
The models may also contain modified versions of the above equations, e.g. to account for non-linearities or optimizable, time-dependent output fractions \eqref{eqn_bottomupmodel_frac}. 

\bigskip
To determine time-dependent CO$_2$ intensities, the what-if analysis analyzes differential changes of the overall CO$_2$ emissions with respect to demand changes. Formally that is,  
\begin{align}
I^{what-if}(co,t) = \frac{\partial M_{tot}}{\partial D(co,t)}.\label{eqn_diff_CO2}
\end{align}
Depending on the model, such derivatives can be computed analytically or via finite differences. 
When computing many CO$_2$ intensities, the finite differences approach can become computationally very burdensome.
Each function evaluation for models like above \eqref{eqn_opt_mod} will be equivalent to a full optimization.
For some problems, however, there is a faster solution as is discussed next.

\section{Efficient What-If analysis for LP models}\label{sec_KKT}

For the special case, where the digital twin is an optimization model in linear programming (LP) form, see \eqref{eqn_opt_mod} above or \citep{SCHABER2012123,osemosys}, we now show how to compute the many partial derivatives \eqref{eqn_diff_CO2} required for the what-if CO$_2$ sensitivities with a single optimization. 

Our exposition is based on standard sensitivity analysis for LPs that can be derived from the Karush-Kuhn-Tucker (KKT) optimality conditions for this type of convex optimization problem \citep{BoyVan04}.
Let the LP be given in standard form as
\begin{align}
\min_{x} \;& c^T x\nonumber\\
s.t. \,&A x \leq b\label{eqn_standard_form}
\end{align}
The problem is convex, it can be dualized into another LP with dual variables $\lambda$, and strong duality holds, i.e. the optima of the original and the dual problem coincide. 
The complementary slackness condition, one part of the necessary KKT conditions, mandates that at the optimum points $x^*,\lambda^*$ of both problems
\begin{align}
(a_i x^{*} - b_i) \lambda_i^{*} = 0,
\end{align}
where $a_i$ is the $i$-th row of $A$. 
Equations for which $\lambda_i > 0$ are called active, since the solution is exactly on the boundary of the feasible region.
Summarizing all the active equations into index set $A$ then allows to write
\begin{align}
A_A x = b_A,
\label{sensanalysis_standard}
\end{align}
where $A_A$, $b_A$ represent the corresponding row subsets.
If under an (infinitesimally) small change of the right hand side $b_A$ the set of active equations stays the same, then this equation also holds for the modified system. It follows that
\begin{align}
A_A \Delta x =  \Delta b_A.\label{sensanalysis_solution}
\end{align}

These general insights about LPs can be employed for energy system models as follows:
in an energy system model the equality condition \eqref{eqn_energy_balance} will always be active and thus be part of index set $A$. 
The CO$_2$ content $M_{tot}$ is one variable in $x$.
To determine a CO$_2$ sensitivity $I^{what-if}(co,t)$, we can therefore solve the linear equation \eqref{sensanalysis_solution} for a right hand side $\Delta b_A$ having only one non-zero entry at the row corresponding to the energy balance for this commodity and time step.
With sparse linear algebra that is applicable to such energy system models, this can be done very efficiently, even for many different energy forms and time slots corresponding to different right hand sides.

\section{Experiments}\label{sec_systems}

To demonstrate the proposed methods, we use an energy system model similar to \eqref{eqn_opt_mod} and simulate two different test energy systems.
The first test energy system, chosen to highlight the effects of renewables' time-variability and of storage, is depicted in \fig{fig:model1}. It has one commodity and five conversion processes and is thus called the single-commodity (SC) model. Its setup and results are discussed in section \ref{sub_sec_SC}.
The second test energy system is chosen to examine the effects of sector coupling. It is depicted in \fig{fig:model2} and has two commodities and five conversion processes. It is called the multi-commodity (MC) model. Its setup and results are presented in section \ref{sub_sec_MC}.
Both systems' parameters are varied in different scenarios to demonstrate specific effects as clearly as possible.

\subsection{Single Commodity (SC) System: Setup and Results}\label{sub_sec_SC}

\begin{figure}[b]
	\centering
	\includegraphics[width=2.5in]{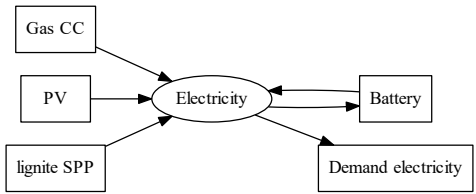}
	\caption{The SC test energy system.
		Rectangles represent a conversion processes and circle shapes the used commodity.}
	\label{fig:model1}
\end{figure}

\begin{table}[!t]
	\renewcommand{\arraystretch}{1.3}
	\caption{Examined scenarios of the SC test energy system)}
	\label{tab_scenario_SCM}
	\centering
	\begin{tabular}{ l | c | c | c | c }
		\hline
		\thead{name of\\scenario}	& \thead{lignite\\SPP} & \thead{gas\\CC} & \thead{PV}&\thead{electric\\storage}\\ 
		\hline\hline
		Scenario 1 			&+	&+	&-	&-	\\\hline
		Scenario 2		&+	&+ 	& + &+ \\\hline 
	\end{tabular}
\end{table}

The SC model is based on the commodity electricity and has five different conversion processes, see \fig{fig:model1}.
The time trajectory of the electric demand is continuously increasing from zero at the start until it reaches its maximum value of $1.5 GW$ at time $10h$.
Electricity is generated by a lignite steam power plant (SPP) with an electric efficiency of $\eta=45\%$, 
a gas combined cycle turbine (CC) with an electric efficiency of $\eta=60\%$,
and a photovoltaic system (PV).
These technologies were chosen to maximize the plausible range of CO$_2$ intensities.
The output capacity of the lignite SPP is $0.75 GW$ ($50\%$ of peak demand), the one of PV is $1.3 GW$ ($92\%$ of peak demand), and gas CC is not limited.
PV availability is additionally time-dependent. 
It starts increasing at time $10h$ towards its maximum at time $17h$, before decreasing back to zero at time $24h$.
The variable cost of electricity production of PV is the lowest, followed by lignite SPP and gas CC, at last.
The CO$_2$ intensity of the consumed natural gas is assumed to be $0.20 t/MWh$, the one of lignite $0.41 t/MWh$\citep{CO2_emission}.

We examine two different scenarios, summarized in Table \ref{tab_scenario_SCM}.
While scenario 1 is as described above, an additional electric storage is considered in scenario 2.
The storage's charging and discharging process are considered lossless, but an hourly self-discharge rate of 1\% per hour applies.

\begin{figure}[p]
	\centering
	\subfloat[Scenario 1]{\includegraphics[scale=0.65]{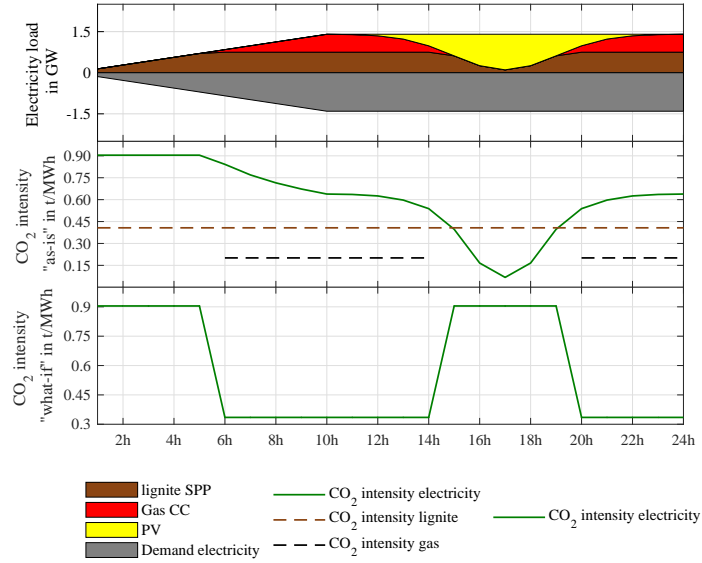}\label{fig:scm_sc1}}\qquad
	\subfloat[Scenario 2]{\includegraphics[scale=0.65]{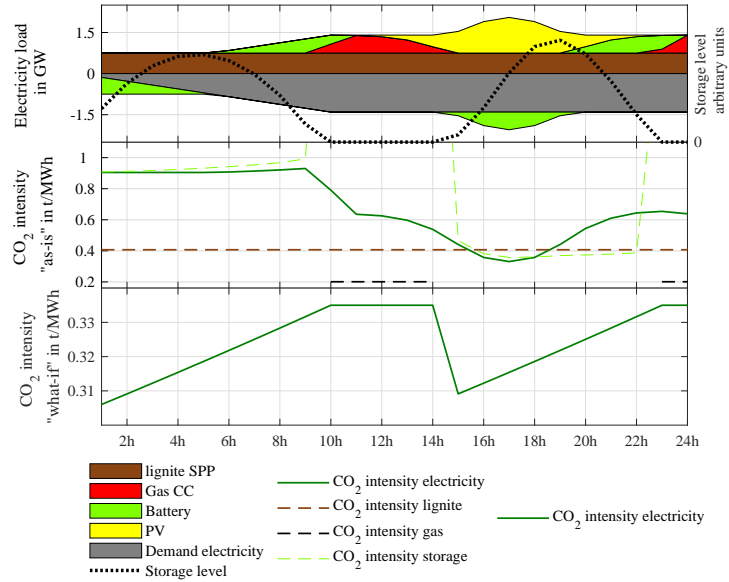}\label{fig:scm_sc2}}

	\caption[CAPTION]{Time trajectories of electricity generation and consumption as well as battery level (top plot), time-dependent CO$_2$ intensities of the as-is analysis (mid) and what-if analysis (bottom) for the different scenarios, see Table \ref{tab_scenario_SCM}. 
		Generation is plotted as positive in the top plot, consumption as negative.
		Note the individual scaling of CO$_2$ intensities in each plot.}
	\label{fig:scm}
\end{figure}

\bigskip
The results of scenario 1 are shown in \fig{fig:scm_sc1}.
Following the merit order of available generation at each point in time, the increasing demand is covered by lignite SPP first and gas CC later, when lignite SPP is at its maximum.
PV replaces both of them, when available.

The computed as-is CO$_2$ intensity of electricity is changing in line with the mix of different sources.
At the beginning, the intensity equals the CO$_2$ intensity of electricity from lignite SPP, that is, $0.9 t/MWh$.
As the gas CC starts operating, the CO$_2$ intensity is decreasing proportionally.
The intensity decrease is even stronger when PV covers a substantial share of the demand. 
The minimum CO$_2$ intensity is reached at the time when the PV has its maximum production.

In contrast to as-is analysis, the CO$_2$ intensity of what-if analysis is not effected by the production mix of the current system.
Instead it examines the CO$_2$ intensity of possible additional generation.
As can be seen in the figure, a completely different behavior of the CO$_2$ intensity follows, with only two levels here.
When lignite SPP is not operating at maximum capacity, the what-if CO$_2$ intensity equals the CO$_2$ intensity of lignite SPP. 
When it is at its limit, the what-if CO$_2$ intensity is the one of gas CC generation, that is, $0.34 t/MWh$.
CO$_2$ neutral electricity is not possible under this view, since all of it is used in the current system already, and none 
is available to cover additional demands.

\bigskip

The results are very different when storage comes into play in scenario 2, see \fig{fig:scm_sc2}.
In this scenario lignite SPP runs at maximum capacity throughout to minimize production cost.
This directly implies higher CO$_2$ emissions overall than in scenario 1.
In the beginning when electricity demand is below SPP production, the storage is charged.
With demand increasing above SPP capacity, the storage is discharged to postpone the operation of costly gas CC.
Gas CC is later displaced by PV. However, not so lignite SPP. Its energy is rather stored to save gas CC operation later,
when PV availability decreases again.

The as-is CO$_2$ intensity of electricity matches the CO$_2$ intensities of the generation mix from gas CC and SPP and from storage.
The CO$_2$ intensity drop for electricity during peak PV availability is smaller than in scenario 1, since lignite SPP continues maximal operation due to the storage option.
The storage content's CO$_2$ intensity depends on the electric CO$_2$ intensity at charging time.
Moreover, it increases with time due to its self-discharge rate.
This is because self-discharge means that storage content is lost but the virtually stored CO$_2$ stays the same.
When the storage is (close to) empty, large but irrelevant storage CO$_2$ intensities result.

The what-if analysis again draws a different picture about the CO$_2$ intensities for this scenario.
Additional use of electricity when gas CC -- the most expensive form of generation -- is already operating, is equal to the CO$_2$ intensity of gas CC, 
since any additional demand would also be covered by this technology.
At times when the storage is discharged to cover demand, one can get additional energy from the storage.
This, however, accelerates the storage's discharge and therefore leads to an earlier restart of gas CC.
That is, if one uses the stored energy early on, gas CC needs to produce more energy later on. 
However, an earlier use of the storage's energy avoids energy losses due to self-discharge.
This means that the use of one unit of energy from storage at time $t_1$ leads to an energy demand from gas CC of less than one at time $t_2$, where $t_1 < t_2$.
The effect increases with the temporal distance between $t_1$ and $t_2$ and explains the slightly increasing CO$_2$ intensities of the what-if analysis during times with non-empty storage tank.

\subsection{Multi Commodity (MC) System: Setup and Results}\label{sub_sec_MC}

\begin{figure}[t]
	\centering
	\includegraphics[width=0.8\linewidth]{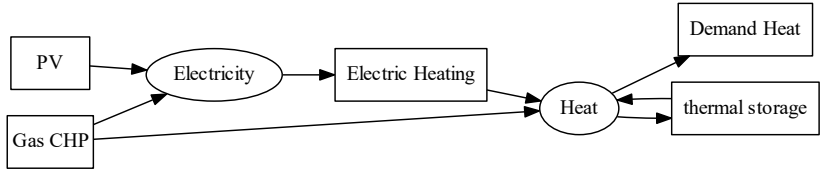}
	\caption{The MC test energy system.
		Rectangles represent a conversion processes and circle shapes the used commodity.}
	\label{fig:model2}
\end{figure}

The second test energy system, depicted in \fig{fig:model2}, has two commodities and five conversion processes.
There is a time-constant electrical demand and a time-varying heat demand with sinusoidal shape over the course of one day. 
A gas driven combined heat and power plant (CHP) supplies electricity and heat with an electric efficiency of $35\%$. The thermal output is flexible and can be ramped up until a $80\%$ total fuel efficiency for electricity and heat is reached.
The CO$_2$ intensity of the gas input is again set as $0.20 t/MWh$.
The modeled heat storage looses $5\%$ of the energy during the charging process, and $1\%$ of the storage content in each hour.
Resistive electric heating can convert electricity into heat with efficiency $\eta=1$. 
The PV plant has a time-dependent availability, simulating one sunny day followed by one without any irradiation. 
Unused PV energy can be curtailed.
The capacity of the PV is set such that during the sunny day the thermal storage can fully be charged via the electric heater. 
The system cost to be minimized is proportional to the amount of consumed gas for the CHP.
PV electricity is assumed to be free.

\begin{table}
	\renewcommand{\arraystretch}{1.3}
	\caption{Examined scenarios of the MC test energy system}
	\label{tab_scenario_MCM}
	\centering
	\begin{tabular}{ l | c | c | c | c | c}
		\hline
		\thead{name of\\scenario}	& \thead{gas\\CHP} & \thead{electric\\heating} & \thead{PV}&\thead{thermal\\storage} & demand\\ 
		\hline\hline
		
		Scenario 1&+&+ & - &+ &\thead{elec: 50\%\\heat: 100\%} \\\hline
		
		Scenario 2&+&+&+&+&\thead{elec: 0\%\\heat: 100\%}\\\hline
		
		Scenario 3&+& - & - & - & \thead{elec: 100\%\\heat: 30\%}\\\hline
	\end{tabular}
\end{table}

\begin{figure}[p]
	\centering
	\subfloat[Scenario 1]{\includegraphics[scale=0.55]{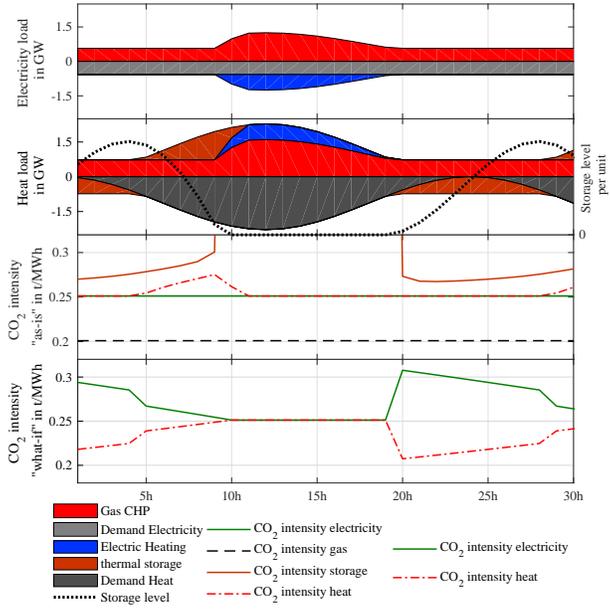}\label{fig:mcm_sc1}}\qquad
	\subfloat[Scenario 2]{\includegraphics[scale=0.55]{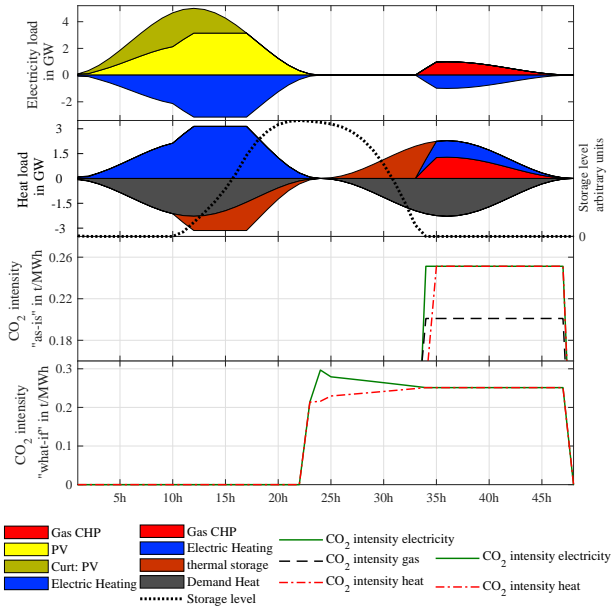}\label{fig:mcm_sc2}}\qquad
	
	\caption[Titel des Bildes]{Time trajectories of generation and consumption, storage levels for electricity (top) and heat (upper mid), CO$_2$ intensities for as-is (lower mid) and what-if analysis (bottom) for the scenarios 1 and 2, see Table \ref{tab_scenario_MCM}. Positive values in the two top plots indicate generation, negative ones consumption.}
	\label{fig:mcm}
\end{figure}

\begin{figure}[t]
	\centering
	\includegraphics[scale=0.55]{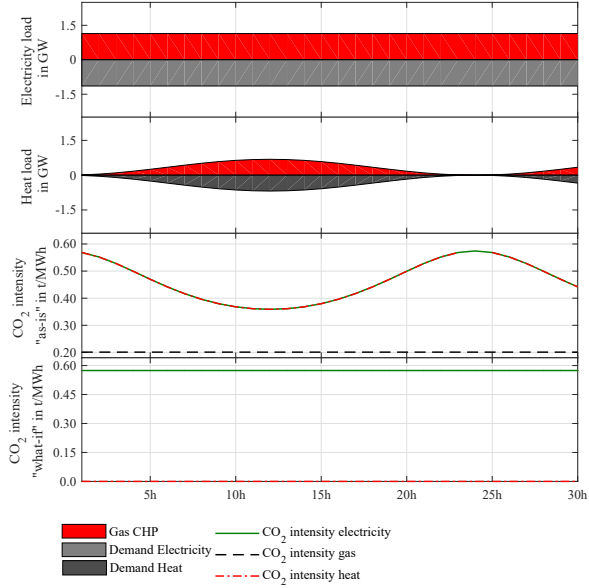}  	   
	\caption[Titel des Bildes]{Time trajectories for generation and consumption and CO$_2$ intensities for as-is (lower mid) and what-if analysis (bottom) for scenario 3, see Table \ref{tab_scenario_MCM}. Positive values in the two top plots indicate generation, negative ones consumption.}
	\label{fig:mcm_sc3} 
\end{figure}

Three different scenarios are examined, as listed in Table~\ref{tab_scenario_MCM}. 
Scenarios 1 models a classic power systems without renewables.
Scenario 2 introduces PV as a fluctuating source, here for heating purposes only.
Scenario 3 demonstrates a special effect when using only the CHP to cover a lower heat, but significantly higher electricity demand.
The simulation results for all scenarios are shown in \fig{fig:mcm} and \fig{fig:mcm_sc3}.

\bigskip
In scenario 1, \fig{fig:mcm_sc1}, the gas CHP is continuously running to cover the electricity and heat demand, always at the operation point with maximum fuel efficiency of $80\%$.
When the electricity demand leads to a heat supply larger than current demand, the thermal storage is charged.
When the heat demand increases, above this supply level, the storage is used.
At the times of peak thermal demand, the CHP produces additional electricity which is converted into heat via the electric heater. 

The optimal operation point of the CHP leads to a constant as-is CO$_2$ intensity of electricity of $0.25 t/MWh = 0.2 t/MWh / 0.8$. 
The same holds for heat, when storage is not discharged. 
The storage CO$_2$ intensity is always above this level due to the charging losses.
As for the SC model, it increases over time due to self-discharge losses.
In total, that means that the as-is CO$_2$ intensity of heat is increasing during times of storage discharge.

The what-if analysis yields equal CO$_2$ intensities of $0.25 t/MWh$ for heat and electricity, when the storage is not operating.
This is a consequence of electric heating which transforms heat to electricity without any losses.
To show and explain the complex interactions in multi-modal system with optimal operation, we now describe in detail the chain of effects, that a electric demand increase would have.
If electricity demand is increased by 1 unit, gas CHP has to use additional 2.86 units of gas to produce this electricity at an efficiency of $35\%$. 
This results in additional 0.57 units of CO$_2$ emissions.
The corresponding heat supply of 1.29 units substitutes the operation of the electric heater, reducing the electric demand by 1.29 units, which in turn reduces the CHP operation by 3.67 gas units and CO$_2$ emissions by 0.73 units.
This, however, would leave 1.65 units of heat demand without supply. 
The gas CHP therefore has to use additional 2.07 units of gas to cover the 1.65 units of heat with 0.93 units of heat supplied directly and 0.72 units of electricity which are transformed to heat via the electric heater ($\eta=1$).
This additional 2.07 units of gas result in 0.41 units of CO$_2$ emission.
The CO$_2$ balance of all these steps is 0.25 units for 1 unit of electricity.

During storage operation the what-if CO$_2$ intensities for electricity and heat are different, 
since electric heating is not operating then.
The what-if CO$_2$ intensity of heat is similar as for the SC model described above.
Electricity's CO$_2$ intensity is mirroring the heat intensity values.
To cover additional electricity demand, gas CHP has to increase production. 
The corresponding heat supply is stored and used later.
However, due to storage losses the savings in later CHP operation for heat are the smaller, the longer the heat stays in the storage.
This explains why the CO$_2$ intensity of electricity is declining over time, unlike the heat intensity.
\bigskip

The results for scenario 2 with renewables are shown in \fig{fig:mcm_sc2} for two days. 
During the first day, all demand can be covered by PV only.
There is also sufficient free energy to fully charge the thermal storage.
Some PV energy even needs to be curtailed due to the limited capacity of the storage. 
During the second day, the heat demand is covered from thermal storage first.
When it is empty, the gas CHP produces electricity and heat to optimally cover the heat demand directly and via electric heating.

As-is analysis yields CO$_2$ intensities of electricity and heat that are zero during day one.
Hence, the storage CO$_2$ intensity is also zero on that day.
During the second day, the CO$_2$ intensity of electricity and heat rises to $0.25 t/MWh$, as a result of the gas CHP operation at the optimal operation point, see scenario 1.
In one time step when the CHP is already running, the provided heat is a mix of stored CO$_2$ neutral heat and CO$_2$ loaded heat produced by gas CHP.
This results in a CO$_2$ intensity of heat of $0.14 t/MWh$ (note that the plot scale is limited at $0.18 t/MWh$ here for optimal presentation).
The scenario shows in comparison with scenario 1 that the storage's contribution to the CO$_2$ intensity of the output energy form depends strongly on the energy mix that was used to fill the storage tank.
It can both lower or increase this value.

The what-if CO$_2$ intensity for both heat and electricity is zero as long as the PV is providing energy and the storage is not fully charged.
Additional demand could then be covered by decreasing the PV curtailments.
When the fully charged state is passed, the what-if CO$_2$ intensities show the same behavior as in scenario 1.

\bigskip
Scenario 3, shown in \fig{fig:mcm_sc3}, has a high electricity demand and a periodic, but low thermal demand.
The CHP has a constant electric operation and the heat demand can be covered directly from the CHP's waste heat at all times.
We chose this scenario to highlight the following possibly counter-intuitive observation of the as-is analysis: 
how much waste heat from the CHP is actually used does not change its operation (typically defined as the mechanical/electrical operation level). 
One could argue that the CO$_2$ intensity of electricity then should also be constant.
As a results of the equal treatment of all output energy forms from one conversion process, the as-is CO$_2$ intensities of both heat and electricity, however, do vary with the combined amount of heat and electricity production.

In the what-if analysis, on the other hand, the CO$_2$ emissions of the CHP operation are fully allocated to electricity, and heat is rated as CO$_2$ free.
This is because CHP operation would have to change in response to electricity demand changes, but not in response to heat demand changes.
Moreover, the what-if CO$_2$ intensity of electricity is time-constant which might seem more natural than the results of as-is analysis here.

\section{Discussion and conclusion}\label{sec_conc}

Our initial target was to develop methods, that allow individual actors in an energy system to understand the CO$_2$ consequences of their local behavior.
This may include analyses of current energy-related behavior as well as studies about the effects of behavioral changes.
Such changes could result both from changed or optimized operation schedules or investments into energy technologies, e.g. new heating systems.
Methods for evaluating individual CO$_2$ footprints could moreover be used for CO$_2$ taxation or pricing, both in energy systems on a nation-wide scale or in local microgrids.

CO$_2$ intensities are a good tool for reducing complex power systems into a single (time-dependent) number with which users can rate their local behavior. 
However, we have demonstrated both theoretically and experimentally, that computing such CO$_2$ intensities is not a trivial task.

\bigskip

Suitable CO$_2$ intensities should fulfill three criteria. 
First, they should correctly mirror the true CO$_2$ consequences.
Second, the method for computing them should be relatively easy and the results should be intuitive. 
On the one hand, this improves users' understanding and acceptance.
On the other hand, the methods need to be easy enough to be explainable in legal text when used for regulatory purposes.
Third, the results should be computable with limited effort both computationally as well as with respect to required data.

Concerning the quality of results we have shown in the experimental section, that an annual average CO$_2$ intensity does not correctly reflect the range of CO$_2$ sensitivities that energy systems with high shares of renewables actually have.
The as-is analysis yields correct results if one wants to analyze current behavior for informatory purposes only.
When considering changes in behavior, the what-if analysis is more relevant.
Since most analysis is only sensible if it leads to actions, we feel that the second argument is slightly more important.

While what-if analysis is easy to explain on a high level, its experimental results are often non-trivial to explain for concrete examples, see e.g. the logical chain of steps needed to explain one value for scenario 1 of the MC model.
Moreover, what-if results might be very sensitive to small changes in the complex energy system.
Consider for example a merit-order model with many plants. Depending on the exact CO$_2$ characteristic of any one of them the CO$_2$ intensity might be different for any current demand value.
As-is analysis generally yields much smoother results since it averages over the whole fleet of processes  / plants generating one commodity.
It also seems more intuitive sometimes, consider for example the low value when the sun is shining in scenario 1 of the SC model.
Unfortunately, sometimes the converse is true, consider e.g. scenario 3 of the MC model.

Concerning ease of computation, 
as-is analysis requires the time-resolved knowledge of all energy flows and all storage levels in the system.
While this data is only partially available today, one could well imagine to collect it in the future. 
The computational burden for solving the set of linear equations is negligible, even for realistically large systems.
For the what-if analysis, a major challenge is the existence of a precise digital twin. 
However, there are many other uses for such models besides CO$_2$ intensities.
This could justify the effort for the twin.
From a computational side, the effort of computing partial derivatives can be reduced with our proposed approach, at least for a large, common class of models.

%

\bigskip

A final endorsement for one of the presented methods is beyond this paper. 
We are looking forwards to see future developments on this topic.
An important next step will be the application of the proposed methodology to real energy systems, e.g. to rate true local projects with these metrics and discuss the consequences with all stakeholders.




\bibliographystyle{elsarticle-num-names} 
\bibliography{references}

%
%
%
\end{document}